\documentclass[english,twocolumn,pra,showpacs]{revtex4}
\usepackage[latin1]{inputenc}
\usepackage{amsmath}
\usepackage{graphicx}
\usepackage[english]{babel}
\newcommand{\beq}{\begin{equation}}
\newcommand{\eeq}{\end{equation}}
\newcommand{\ket}[1]{| #1 \rangle}

\usepackage{babel}
\begin{document}

\title{Universal optimal broadband photon cloning and entanglement creation in one dimensional atoms}

\author{D. Valente$^{1}$}\email{valente.daniel@gmail.com}
\author{Y. Li$^{2}$}
\author{J. P. Poizat$^{1}$}
\author{J. M. G\'erard$^{3}$}
\author{L. C. Kwek$^{2}$}
\author{M. F. Santos $^{4}$}
\author{A. Auff\`eves$^{1}$}

%\singlespacing

\affiliation{$^{1}$ Institut N\'eel-CNRS, Grenoble, France}

\affiliation{$^{2}$ Centre for Quantum Technologies, National University of Singapore}

\affiliation{$^{3}$ CEA/INAC/SP2M, Grenoble, France}

\affiliation{$^{4}$ Departamento de F\'isica, Universidade Federal de Minas Gerais, Belo Horizonte, Brazil}

\begin{abstract}
We study an initially inverted three-level atom in the lambda configuration embedded in a waveguide, interacting with a propagating single-photon pulse. Depending on the temporal shape of the pulse, the system behaves either as an optimal universal cloning machine, or as a highly efficient deterministic source of maximally entangled photon pairs. This quantum transistor operates over a wide range of frequencies, and can be implemented with today's solid-state technologies.
\end{abstract}

\maketitle

\section{Introduction} Perfect cloning of a quantum state is forbidden by the linearity of quantum mechanics \cite{Wooters},
otherwise, it could be exploited for superluminal communication \cite{Gisin}.
Nevertheless, imperfect cloning is possible, and optimal fidelities have been computed \cite{vlado}, which has interesting applications in the framework of quantum cryptography \cite{valerio} and state estimation \cite{artur}.
On the other hand, entanglement is a fundamental resource in quantum mechanics, lying at the heart of efficient quantum computation algorithms and quantum communication protocols.
Here we present a versatile device that can be operated either as a universal cloning machine, or as a deterministic source of EPR pairs, the functionality being chosen
by the spectral shape of the signal photon wavepacket. This quantum transistor, working at the single photon level, relies on a particular ``one-dimensional (1D) atom'' \cite{1D},  made of a three-level atom in the lambda configuration, embedded in a one-dimensional electromagnetic environment. Unlike more common 1D atoms made of a single atom in a leaky cavity, our system is broadband, can operate over a wide range of frequencies, and processes propagating single photon pulses with negligible input/output coupling losses.

Since the pioneering work of Kimble and coworkers \cite{1D}, 1D atoms have been the subject of numerous experimental and theoretical investigations due to their potential in quantum communication and information processing. In particular, they provide
optical non-linearities at the single photon level \cite{Chang, GNL, 1d}, paving the road towards the implementation of efficient photonic gates \cite{Kojima_gate}. At the same time, light emitted by the atom
interferes with the pump, leading to interesting phenomena like dipole induced reflection \cite{GNL} or super-bunching in the transmitted light \cite{Chang}.
First held with two-level systems, the study of one-dimensional atoms now tackles more complex structures such as three-level atoms in the V shape, $\Lambda$ shape or ladder configuration, thus
opening the path to the exploration of other promising effects such as single photon transistor \cite{Chang},  electromagnetically induced transparency \cite{Roy, Sorensen}, and full quantum computation \cite{marcelo1, marcelo2}.
These level schemes eventually involve transitions of different frequencies, where the broadband behavior of the 1D environment is of utmost importance.
From the experimental perspective, 1D atoms can be realized with semi-conducting systems, like a quantum dot embedded in a photonic wire.
This device has been probed as a highly
efficient semi-conducting single photon source \cite{Claudon}. Lambda configuration for the emitter can be obtained, taking advantage of the
two possible biexcitonic transitions in quantum dots \cite{lambdaQD} or the different spin states in the optical transitions in a single N-V center \cite{NV}, for instance.
Superconducting qubits in circuit QED offer another natural playground for the exploration of 1D atoms properties \cite{olivier1,olivier2}.
As a matter of fact, EIT \cite{Abdumalikov}, single photon routing \cite{Hoi}, and ultimate amplification \cite{Astafiev} have been demonstrated, building on the three-level structure of transmons or superconducting loops efficiently coupled to microwave sources of two different frequencies.

\section{Stimulating a lambda 1D-atom with a single photon}
\begin{figure}[h!]
\begin{center}
\includegraphics[width=1\linewidth]{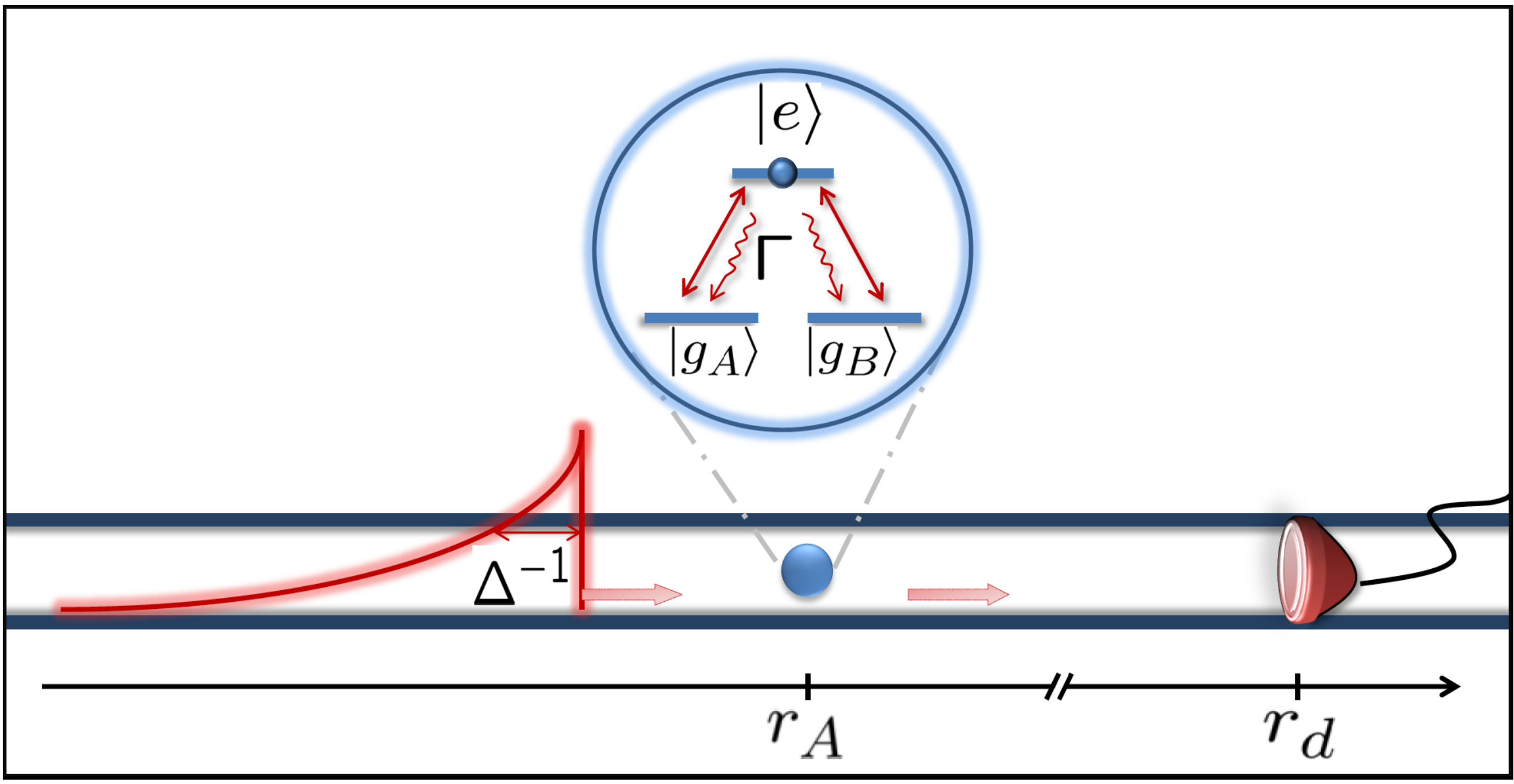}
\caption{Scheme of the 1D atom in lambda configuration with incoming photon of arbitrary polarization and exponential wavepacket shape. In the model, the atom at position $r_A$ is embedded in a semi-infinite 1D electromagnetic channel, so that the emitted light propagates only in the forward direction and is detected at $r_d$ arbitrarily far from the emitter.}
\label{fig:scheme}
\end{center}
\end{figure}

Here we study an initially inverted atom in the lambda configuration interacting with a one-dimensional electromagnetic environment, as pictured in Fig. \ref{fig:scheme}.
At the initial time, a single photon is sent to the atom and eventually stimulates the atomic emission, a situation reminiscent of that in Ref. \cite{1d}, the study here being performed for a quantized incident field as in \cite{daniel2}.
The shape of the wave packet is chosen to be exponential, which corresponds to the spontaneous emission by another neighboring atom.
The two atomic transitions are supposedly degenerated, respectively coupled with the same strength to two electromagnetic continua of orthogonal polarizations denoted $a_\nu$ and $b_\nu$.

\begin{figure}[h!]
\begin{center}
\includegraphics[width=1\linewidth]{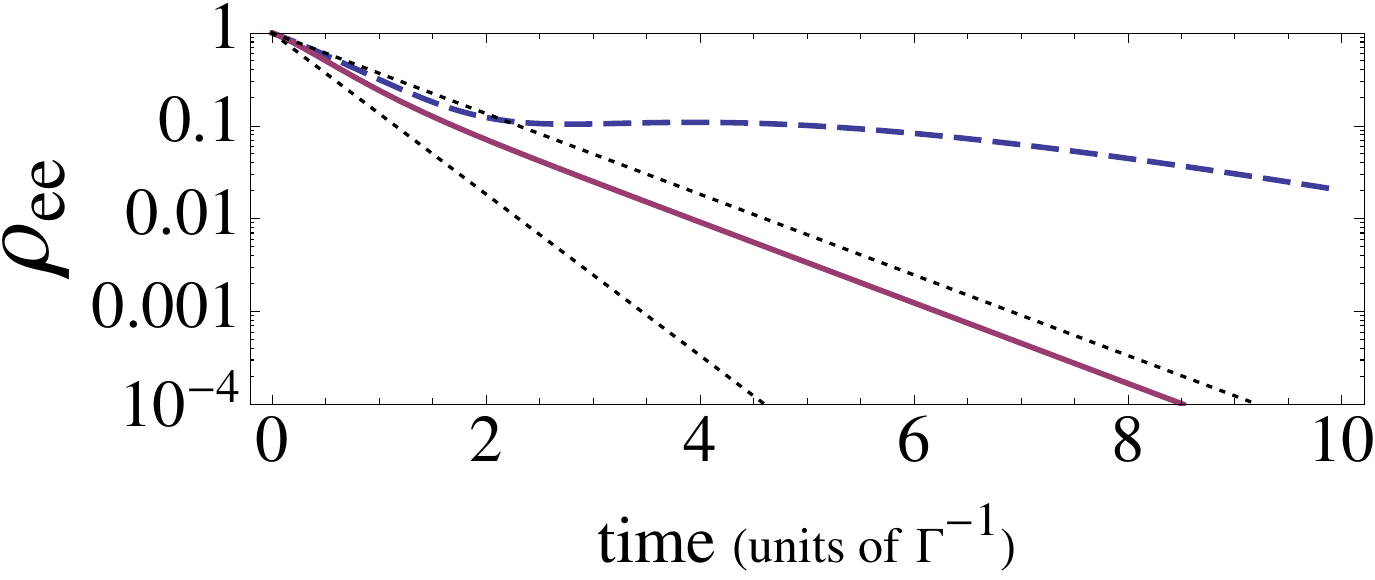}
\caption{Excited state population as a function of time for different spectral widths $\Delta = 0.5\Gamma$ (dashed curve) and
$\Delta = 2\Gamma$ (solid curve). The upper dotted curve is the spontaneous emission exponential decay, for reference.
The lower dotted one is the stimulated emission upper bound, i.e., $\exp{(-2\Gamma t)}$.}
\label{fig:population}
\end{center}
\end{figure}

We consider the case where the continuum of modes has only one direction of propagation, so that the atom can only emit light in one direction as in \cite{Kojima_gate}. This semi-infinite waveguide model could correspond, in principle, to a physical situation where a mirror \cite{friedler}, or a metallic nanotip \cite{Sorensen}, is placed close to the atom, just to mention potential realizations. 
This is valid as long as the distance between the emitter and the mirror is smaller than the coherence length of the field. 
The interaction Hamiltonian of the system is
\beq
H_\mathrm{I} = \sum_\nu -i\hbar g_\nu \left[a_\nu \sigma^a_{+} + b_\nu \sigma^b_{+} - \mbox{H.c.}\right],
\eeq
where
$\sigma^{a}_{+} = |e\rangle \langle g_{A}|$ is the atomic creation operator from the ground state $a$, and analogously for $\sigma^b_{+}$.
Note that the problem is totally symmetrical with respect to any change of polarization basis, so that we can choose an arbitrary polarization $a$ for the
incident photon, without restricting the generality of the problem. The state of the atom-field system at the initial time can be written
$|\psi(0)\rangle = \sum_\nu \psi^a_\nu(0)\ a^\dagger_\nu |e, 0\rangle$, where in the spatial representation with coordinate $r$ we have $\psi^a(r,0) \equiv \sum_\nu \psi^a_\nu(0)\ e^{ik_\nu r}
= \mathcal{N} \Theta(-r)\exp{\left(\frac{\Delta}{2}+i\nu_L\right)\frac{r}{c}}$, and $c$ is the speed of light.
We denote $\Delta$ as the spectral width of the wave packet and $\delta = \nu_L - \nu_A$ its detuning with respect to the atomic frequency $\nu_A$.
The normalization is $\mathcal{N}^2 = 2\pi \rho_{\mathrm{1D}} \Delta$, where $\rho_{\mathrm{1D}}$ is the 1D density of modes ($\sum_\nu \rightarrow \int d\nu \rho_{\mathrm{1D}}$) and $\Theta(x)$ denotes the Heaviside step function.
The dynamics is obtained by analytically solving the Schr\"odinger equation using the ansatz
\begin{eqnarray}
|\psi(t)\rangle &=&\sum_\nu [\psi^a_\nu(t)\ a_\nu^\dagger + \psi^b_\nu(t)\ b_\nu^\dagger]|e,0\rangle +\nonumber \\
&&\sum_{\nu_1,\nu_2} [\phi^{aa}_{\nu_1,\nu_2}\ a_{\nu_1}^\dagger a_{\nu_2}^\dagger +
 2\phi^{ab}_{A\ \nu_1, \nu_2}\ a_{\nu_1}^\dagger b_{\nu_2}^\dagger]|g_A, 0\rangle + \nonumber\\
&&[\phi^{bb}_{\nu_1,\nu_2}\ b_{\nu_1}^\dagger b_{\nu_2}^\dagger +
2 \phi^{ab}_{B\ \nu_1, \nu_2}\ a_{\nu_1}^\dagger b_{\nu_2}^\dagger]|g_B, 0\rangle,
\label{pop}
\end{eqnarray}
 for the state. We have solved a self-consistent differential equation for the probability amplitudes $\psi^{a(b)}(r,t)$ from which
we could also find the solutions for $\phi^{aa(ab)}(r_1,r_2,t)$, as shown below.
Both excited-state amplitudes satisfy 
\begin{eqnarray}
&\left[\frac{\partial}{\partial t} + c\frac{\partial}{\partial r}\right]&\psi^{a,b}(r,t)=  - \left(\frac{\Gamma}{2}+i\nu_A\right)\psi^{a,b}(r,t)\nonumber\\
&&- \frac{\Gamma}{2}  \Theta(r)\Theta(t - r/c) \psi^{a,b}(-r,t - r/c),
\end{eqnarray}
for which the solution reads 
\begin{eqnarray}
\psi^{a,b}(r,t)&=&\psi^{a,b}(r - ct,0)e^{-\left(\frac{\Gamma}{2}+i\nu_A\right)t}\nonumber\\
&& -(\Gamma/2)\Theta(r)\Theta(t - r/c)e^{-(\frac{\Gamma}{2}+i\nu_A)t}\nonumber\\
&&\times e^{-\left(\frac{\Gamma}{2}+ i \nu_A\right)(t - r/c)}  \nonumber\\
&& \int_{t -  r/c}^t  e^{\left(\frac{\Gamma}{2}+i\nu_A\right)t'}\psi^{a,b}(-ct',0)\ dt'.
\end{eqnarray} 
This allows us to compute the two-photon amplitudes, which read 
\begin{eqnarray}
\phi^{aa}(r_1,r_2,t) &=& \frac{\sqrt{\pi \rho \Gamma} }{2} \nonumber\\
&& \times [\Theta(t-r_2/c)\Theta(r_2)\psi^{a}(r_1-r_2,t-r_2/c)+\nonumber\\
&&\Theta(t-r_1/c)\Theta(r_1)\psi^{a}(r_2-r_1,t-r_1/c)]
\end{eqnarray}, 
\begin{eqnarray}
\phi^{ab}_A(r_1,r_2,t) &=&  \frac{\sqrt{\pi \rho \Gamma}}{2}\Theta(t-r_1/c)\Theta(r_1)\nonumber\\
&& \times \psi^{b}(r_2-r_1,t-r_1/c)
\end{eqnarray}
and 
\begin{eqnarray}
\phi^{ab}_B(r_1,r_2,t) &=& \frac{\sqrt{\pi \rho \Gamma}}{2} \Theta(t-r_2/c)\Theta(r_2)\nonumber\\
&& \times \psi^{a}(r_1-r_2,t-r_2/c).
\end{eqnarray}
As the problem is Hamiltonian the number of excitations is conserved during the evolution and is fixed to $2$.
The functions $\psi^{a(b)}(r,t)$ give direct access to the evolution of the atomic excited-state population
$\rho_{\mathrm{ee}} = \langle e|\mbox{Tr}_{\mathrm{field}}[|\psi(t)\rangle\langle \psi(t)|]|e\rangle$  %\cite{supplemental}
which is plotted in Fig. \ref{fig:population}.

Because of its coupling to a continuum, the
atom irreversibly relaxes towards one of the ground states by emitting a photon. The typical rate for the relaxation is given by
$\Gamma = \sum_\nu 4\pi g_\nu^2\delta(\nu-\nu_A)$,
which is the spontaneous rate derived from the Wigner-Weisskopf approach.
Note that the expression for $\Gamma$ takes into account the presence of the mirror in the semi-infinite waveguide. In the full transmitting/reflecting waveguide, the spontaneous decay rate would be given by $\Gamma_{\mathrm{full}} = 2\Gamma$. For experimental purposes, this rate can be measured independently and its actual value does not affect our analysis.
Depending on the adimensional width of the wavepacket $\Delta/\Gamma$,
the emission of the photon is more or less efficiently stimulated. The dashed curve in Fig. \ref{fig:population} shows a reabsorption feature
at $\Delta = 0.5  \Gamma$, for instance. Contrary to intuition, the optimal stimulation does not occur for the mode matching with spontaneous emission ($\Delta = \Gamma$). In the configuration here analyzed, the most efficient stimulation is reached for $\Delta=2\Gamma$, as shown by the solid curve in Fig. \ref{fig:population}. In this case the atom relaxes almost $1.5$ times faster than in the spontaneous emission case.
The maximal rate one can expect by stimulating with a single photon is twice the spontaneous emission rate,
which can be obtained with a two-level atom in the same waveguide configuration used in this paper \cite{daniel2.2}.
In the limiting cases where $\Delta\gg \Gamma$ and $\Delta\ll \Gamma$ corresponding to a wavepacket respectively localized in the time domain or the frequency domain, the overlap with the atomic mode is negligible and we are brought back to the spontaneous emission behavior.

\section{Universal optimal cloning}
\begin{figure}[h!]
\begin{center}
\includegraphics[width=0.95\linewidth]{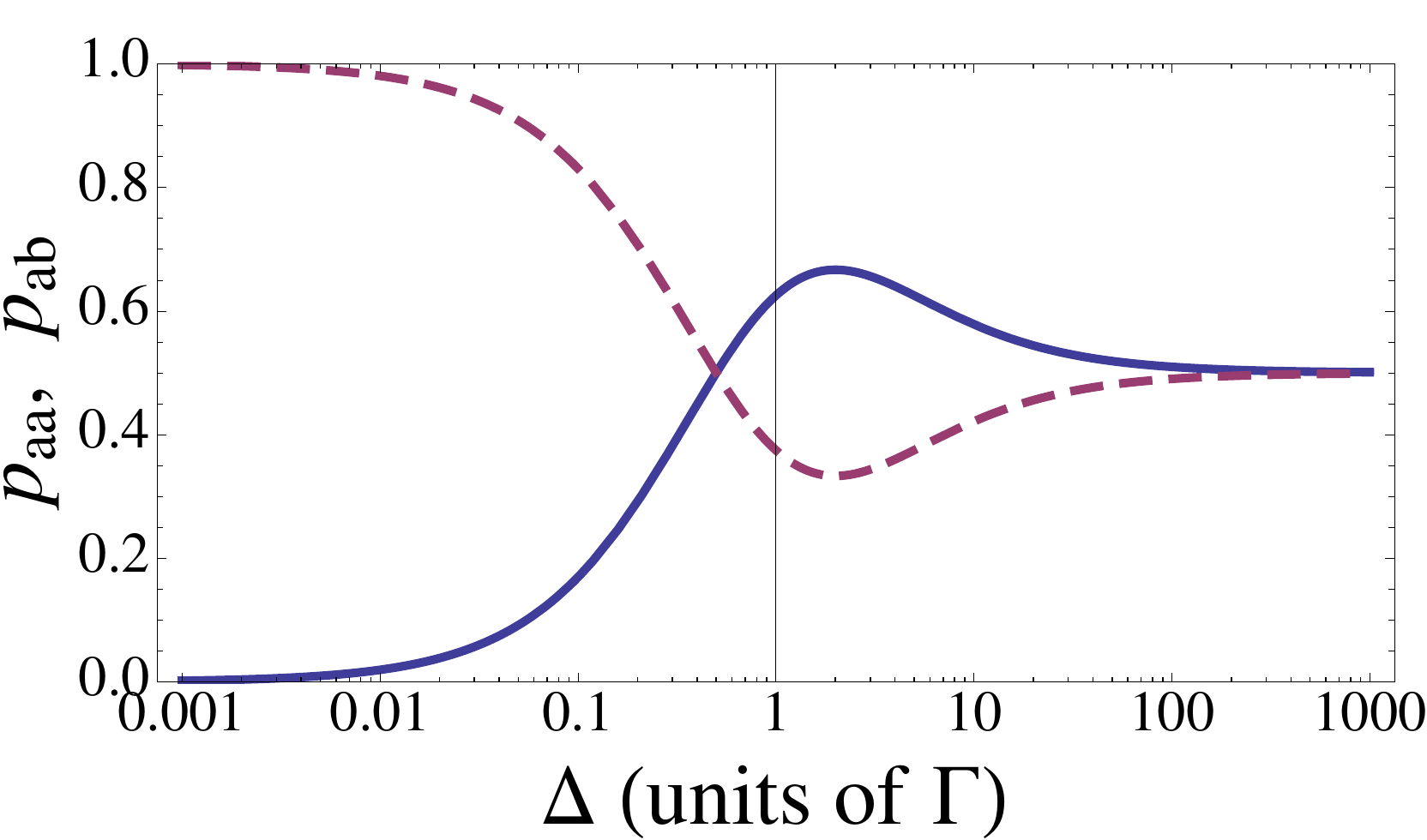}
\caption{Probabilities $p_{aa}$ (solid blue curve) and $p_{ab}$ (dashed red curve) for
two photons created with the same polarizations and orthogonal polarizations, respectively.}
\label{fig:paa.pab}
\end{center}
\end{figure}

In addition to fast atomic relaxation, the other feature of stimulated emission is the likelihood of the atom emitting a photon in the stimulating mode.
This property is quantified by the probabilities $p_{aa}$ and $p_{ab}$ to produce the two photons with the same polarization or with two distinct polarizations respectively, in the end of the relaxation process. We have
\beq
p_{aa} = \frac{\Delta (4\Gamma+\Delta)}{2(\Gamma + \Delta)^2},
\ \mbox{and} \
p_{ab} = \frac{1}{2} \left(1+\frac{\Gamma^2-2\Delta\Gamma}{(\Gamma+\Delta)^2}\right),
\eeq
given our choice for the initial state (note that $p_{bb}=0$).
These quantities are obtained from $ p_{aa} = \sum_{\nu,\nu'}2|\phi^{aa}_{\nu,\nu'}(t)|^2$ and $p_{ab} = \sum_{i=A,B}\sum_{\nu,\nu'}4|\phi^{ab}_{i,\nu,\nu'}(t)|^2$ %\cite{supplemental}
taken for $\Gamma t\rightarrow \infty$ and are plotted in Fig.\ref{fig:paa.pab} with respect to the parameter $\Delta/\Gamma$.
When $\Delta\gg \Gamma$ (highly localized wavepacket in time), spontaneous emission takes place, hence the probabilities for the atom to emit in the modes $a$ or $b$ are equal and $p_{aa} = p_{ab}=1/2$. As previously stated, maximal stimulation occurs for a packet that is shorter than the spontaneous emission shape.
When $\Delta=2\Gamma$, where atomic emission is the most efficiently stimulated, we have $p_{aa} = 2/3$. This value is optimal;
in this point indeed, the atomic emission
in the stimulating mode $a$ is twice more probable than in the empty mode $b$, which is the maximum ratio one can expect when the stimulating mode contains a single photon.
So far such a ratio has only been evidenced in cavities \cite{haroche} where the effect of bosonic amplification naturally arises, the price to pay being the reversibility of stimulated emission.
Oddly enough, this ratio is preserved here where the atomic emission is stimulated in a continuous distribution of modes, hence irreversible.
This precise relation $p_{aa} = 2/3$ and $p_{ab} = 1/3$ also corresponds to the
maximal fidelity
$\mathcal{F} = p_{aa} \mathcal{F}_{\mathrm{right}}+p_{ab} \mathcal{F}_{\mathrm{wrong}}
= \frac{2}{3}\times 1+\frac{1}{3}\times\frac{1}{2} = \frac{5}{6}$
one can reach in cloning the incident photon polarization \cite{vlado,CS,Antia}.
Since, as previously stated, the interaction Hamiltonian is invariant under unitary transformations of the polarization basis, this device can indeed be operated as a universal optimal cloning machine.
Exploiting stimulated emission to clone a quantum state has
inspired proposals where three level atoms coupled to cavities were used as cloners, and optimal cloning was also theoretically demonstrated \cite{CS,Kempe,Zou}.
The use of a high-quality cavity implies a confination of the photons,
which brings the drawback of reducing the deterministic access to the clones.
Furthermore, the present effect could not be obtained in a dissipative cavity. In that case, the atomic excitation
would escape from the cavity in a typical time $1/\kappa$, much faster than the stimulation time scale of $\kappa/g^2$,
where the atom-cavity coupling strength $g$ satisfies $g \ll \kappa$ in the weak coupling regime.
By contrast, optimal cloning in a one-dimensional environment can be implemented by exploring the pulse shape of the photons,
building on the broadband coupling of the emitter with the light field. Hence propagating fields can be cloned, a highly desirable property for all practical purposes \cite{valerio}. Further details on the difference between genuine broadband dynamics and leaky cavities are found in Ref.\cite{leakycav}.

\section{Deterministic entanglement production}
The case where $\Delta\ll\Gamma$ corresponds to a monochromatic (long) incident photon. In this situation, the probabilities become $p_{aa}\rightarrow 0$ and $p_{ab}\rightarrow1$
as shown in Fig.(\ref{fig:paa.pab}). Even though this case corresponds to spontaneous emission, as in the $\Delta\gg \Gamma$ case, the characteristics of the light are drastically different. In particular, one never gets two photons of the same polarization. This effect can be understood by noting that a monochromatic photon of polarization $a$ impinging on a lambda atom prepared in state $g_a$ is entirely scattered in mode $b$, as shown below, leading to the mapping $\ket{a_L, g_a}\rightarrow -\ket{b_L,g_b}$. The subscript $_L$ describes a long wavepacket. The shape of the wavepacket is conserved during such scattering process.
The semi-infinite geometry (which takes the mirror into account) is a necessary condition for this state transfer to happen, as it provides the proper interference conditions. This can be shown by means of the outgoing photon wavepackets $\phi^a(r,t)$ and $\phi^b(r,t)$ derived from the initial state $\ket{g_A}\ket{a}$ (single-excitation subspace), which read
\begin{eqnarray}
\phi^{a,b}(r,t) &=& \phi^{a,b}(r-ct,0)\nonumber\\
&&+\sqrt{\Gamma\pi\rho_{\mathrm{1D}}}\Theta(r)\Theta(t-r/c)\psi(t-r/c).
\end{eqnarray}
The excited-state amplitude in this case is given by $\psi(t) = - \sqrt{\frac{\Gamma}{\pi\rho_{1D}}}\mathcal{N} e^{-\left(\frac{\Gamma}{2}+i\nu_A \right) t}
\left( \frac{e^{(\frac{\Gamma-\Delta}{2}-i\delta_L) t}-1}{\Gamma-\Delta-2i\delta_L}\right)$, which in the $\Delta \ll \Gamma$ (long wavepacket) limit becomes 
\beq
\Theta(t-r/c)\ \psi(t-r/c)\approx - \frac{1}{\sqrt{\Gamma\pi \rho_{1D}}}\ \phi^{a}(r-ct,0), 
\eeq
where $\phi^a(r,0) = \mathcal{N} \Theta(-r)\exp{\left(\frac{\Delta}{2}+i\nu_L\right)\frac{r}{c}}$.
The $\pi$-phase shift in $\psi(t-r/c)$ creates an exact destructive interference that cancels the amplitude for polarization $a$, $\phi^a(r,t)=0$. Were it a full waveguide, the amplitude created from the interaction,  namely, $\sqrt{\Gamma\pi\rho_{\mathrm{1D}}}\Theta(r)\Theta(t-r/c)\psi(t-r/c)$, would symmetrically split itself through both reflection and transmission channels, preventing completely destructive interference. For the amplitude of polarization $b$, no intereference takes place since it is initially in vacuum state $\phi^b(r,0) = 0$, so $\phi^b(r,t) = -\phi^a(r-ct,0)$. Hence the initial shape of the wavepacket is conserved during the map $\ket{a_L, g_a}\rightarrow -\ket{b_L,g_b}$.
A related effect is found in Ref.\cite{statetransfer}. 

The succession of steps is basically the following. First, the atom spontaneously
emits a photon with equal probability in mode $a$ or $b$, ending up respectively in the ground state $g_a$ or $g_b$. At this point the atom and the field are entangled in a global state
that can schematically be written $(1/\sqrt{2})\ket{a_L}(\ket{g_b,b_S}+\ket{g_a,a_S})$. The index $_S$ labels the short wavepacket obtained through the spontaneous emission process.
The atom interacts with the incoming photon $\ket{a_L}$ if it is in the state $\ket{g_a}$, otherwise it is transparent.
In any case, it finally decouples and the entanglement is entirely mapped on the light field, the final two-photon state being
\beq
\ket{\mbox{final two photons}} = \frac{1}{\sqrt{2}} \ (\ket{b_S,a_L} - \ket{a_S,b_L}). 
\eeq
Note that the two photons are completely distinguishable in that state ($\langle a_S|a_L\rangle = \langle b_S|b_L\rangle = 0$), given that the short one lies within
the lifetime of the atom and the long one extends over a thousand lifetimes or more, and hence they can be separated in practice.
In this operating point, the device acts as a deterministic source of EPR pairs, triggered by a single pump photon.
In this process, EPR pairs can thus be produced efficiently over a wide range of frequencies, offering a promising alternative to other protocols based on parametric down conversion \cite{parametric} or biexcitonic radiative cascade \cite{jpp1}.

\section{Possible error sources}
In a realistic scenario, two noise sources must be taken into account, namely, the decay rate into the environmental 3D channels $\gamma$ and
the pure dephasing rate $\gamma^*$ present in solid-state systems.
The former is usually quantified by the parameter $\beta = \Gamma/(\Gamma+\gamma)$ which can reach $0.98$ in 1D nanophotonic systems made of photonic wires \cite{Claudon} or 1D waveguides in photonic crystals \cite{photonicCrystal1,photonicCrystal2}, and almost $1$  in circuit QED \cite{Abdumalikov}.
Pure dephasing rates of $\gamma^* \approx 0.1 \Gamma$ have been measured in quantum dots \cite{dephasingQD} and superconducting qubits \cite{astafievPureDephasing}.
From Ref.\cite{1d}, we could estimate that such imperfections would affect the cloning fidelity and the entanglement
by a factor of the order $\sim \beta (1-\gamma^*/\Gamma)$, for $\beta \approx 1$ and $\gamma^* \ll \Gamma$.
This would lower the real cloning fidelity and entanglement to about $90\%$ of their optimal values for circuit QED systems
and $88\%$ for nanophotonic systems.
In addition to building cleaner systems, dynamical decoupling approaches have been proposed to reduce dephasing in lambda-type systems \cite{plenio}.

\section{Conclusions}
In conclusion, we have presented a versatile device that can realize either universal optimal cloning
or maximal entanglement in photon polarization, depending only on the spectral shape of the incoming photon. A single three-level atom in a 1D open space has been used, giving rise to a genuine broadband system. A realistic single-photon pulse shape has been considered, yielding maximal efficiencies on both processes. The photonic propagation makes the reported effects especially attractive as far as realistic implementations of quantum information processing are concerned.

\section*{Acknowledgements}
This work was supported by the Nanosciences Foundation
of Grenoble, the CNPq and Fapemig from Brazil, the ANR projects ``WIFO'' and ``CAFE'' from France, and the Centre for Quantum
Technologies in Singapore.

\end{document}